\documentclass[twocolumn,english,prl,superscriptaddress,showpacs]{revtex4}
\usepackage[T1]{fontenc}
\usepackage[latin9]{inputenc}
\usepackage{graphicx}
\usepackage{color}
\usepackage{ulem}

\makeatletter

\@ifundefined{textcolor}{}
{%
 \definecolor{BLACK}{gray}{0}
 \definecolor{WHITE}{gray}{1}
 \definecolor{RED}{rgb}{1,0,0}
 \definecolor{GREEN}{rgb}{0,1,0}
 \definecolor{BLUE}{rgb}{0,0,1}
 \definecolor{CYAN}{cmyk}{1,0,0,0}
 \definecolor{MAGENTA}{cmyk}{0,1,0,0}
 \definecolor{YELLOW}{cmyk}{0,0,1,0}
 }

\makeatother

\newcommand{\CMS}{CoMnSi}
\newcommand{\CNMS}{Co$_{0.95}$Ni$_{0.05}$MnSi}

\newcommand{\Tt}{$T_{t}$}
\newcommand{\done}{$d_{\rm{1}}$}
\newcommand{\dtwo}{$d_{\rm{2}}$}

\usepackage{babel}

\begin{document}

\title{Giant magneto-elastic coupling in a metallic helical metamagnet }
\author{A. Barcza}
\affiliation{Dept. of Materials Science and Metallurgy, University of 
Cambridge, New Museums Site, Pembroke Street, Cambridge, CB2 3QZ, UK}
\author{Z. Gercsi}
\affiliation{Dept. of Materials Science and Metallurgy, University of 
Cambridge, New Museums Site, Pembroke Street, Cambridge, CB2 3QZ, UK}
\affiliation{Dept. of Physics, Blackett Laboratory, Imperial College London, London SW7 2AZ UK}
\author{K.S. Knight}
\affiliation{ISIS, Rutherford Appleton Laboratory, Oxon OX11 0QX, UK}
\author{K.G. Sandeman}
\affiliation{Dept. of Physics, Blackett Laboratory, Imperial College London, London SW7 2AZ UK}

\begin{abstract}
Using high resolution neutron diffraction and capacitance dilatometry we show that the thermal evolution of the helimagnetic state in CoMnSi is accompanied by a change in inter-atomic distances of up to 2\%, the largest ever found in a metallic magnet.  Our results and the picture of competing exchange and strongly anisotropic thermal expansion that we use to understand them sheds light on a new mechanism for large magnetoelastic effects that does not require large spin-orbit coupling.
\end{abstract}
\maketitle
Most materials change shape in a magnetic field.  Usually the effect, and the so-called magneto-elastic interaction from which it derives are small, especially away from a phase transition.  However, large magneto-elastic interactions are crucial to a range of new, technological materials in which multiple order parameters exist simultaneously and are coupled.  These include ferromagnetic shape memory materials~\cite{vasiliev_1999a}, multiferroics~\cite{lee_2008a}, and magnetic refrigerant materials~\cite{tegus_2002a}.  Despite their relevance, atomically-resolved observations of large magneto-elastic effects within a given crystal structure are very rare.  

Giant magneto-elastic coupling was recently observed in multiferroic hexagonal manganites~\cite{lee_2008a}, where it is two orders of magnitude larger than in any other magnetic material.  In those compounds it is believed to be central to magneto-electric coupling and multi-ferroicity and is thought to arise, unusually, from strongly varying exchange interactions where the co-ordination of Mn atoms in MnO$_5$ trigonal bipyramids removes the orbital degeneracy and Jahn-Teller mechanism typically found in MnO$_6$-derived structures.  Here we show that exchange-derived giant magneto-elastic interactions are not limited to multiferroic oxides, but may be much more general, if one examines materials that possess competing exchange interactions relieved by temperature or applied field. The system we study is CoMnSi, a metallic antiferromagnet.

We previously observed a large MCE in CoMnSi, a metamagnet that has a field- and temperature-induced transition from a low temperature, non-collinear incommensurate helical antiferromagnetic (AFM) state to a high magnetisation state~\cite{sandeman_2006a}.  CoMnSi is structurally similar to MnP, a system in which the field-dependence of non-collinear magnetic states has been well studied ~\cite{reis_2008a} and in which a rare Lifshitz tricritical point is seen~\cite{becerra_1980a}.  However the similarity of the magnetic structures in MnP and in CoMnSi is not so well known, due to a lack of single crystals and of temperature-dependent neutron diffraction data.  Here we focus on the iso-structural evolution of the crystal lattice within the helical groundstate of CoMnSi.  We show that, within the AFM state and well below the zero field N\'{e}el temperature there is a giant, and opposing, change in the two shortest Mn-Mn distances, of around 2\%.  This change has two consequences.  Firstly it brings about an Invar-like effect in sample volume in zero magnetic field.  Secondly, it couples strongly to the suppresion of helimagnetism  in finite magnetic fields and brings about a tricritical point, with enhanced magnetocaloric and magnetostrictive effects.

The samples that we study here were formed by co-melting appropriate amounts of high purity elements under argon in an induction furnace, followed by post-annealing and slow cooling.  Details are given elsewhere~\cite{sandeman_2006a}. 
Magnetic measurements in fields of up to 9~Tesla were performed on polycrystalline samples with a Cryogenic Ltd. vibrating sample magnetometer (VSM). Structural characterisation was carried out in up to three ways.  Firstly, we conducted a Rietveld refinement of data from room temperature X-ray difraction using Cu-K$\alpha$ radiation. Secondly, thermal expansion and magnetostriction were measured with a miniature capacitance dilatometer~\cite{rotter_1998a} using polycrystalline pieces cut with a wire saw to have at least one flat face. The Cryogenic Ltd. superconducting magnet and cryostat were used in this experiment.  
Thirdly, on two samples neutron diffraction was carried out at the time-of-flight High Resolution Powder Diffractometer (HRPD) at ISIS, UK. This instrument has a resolution of  $\Delta d/d \sim 1\times10^{-4}$ and was used at temperatures between 4.2~K and 500~K.  X-ray diffraction at room temperature showed that all samples were single phase to within experimental resolution.  Structural refinements of these data agreed well with neutron diffraction data results close to room temperature.  We therefore only discuss unit cell refinements where obtained by high resolution neutron diffraction.

\begin{figure}
\includegraphics[width=\columnwidth]{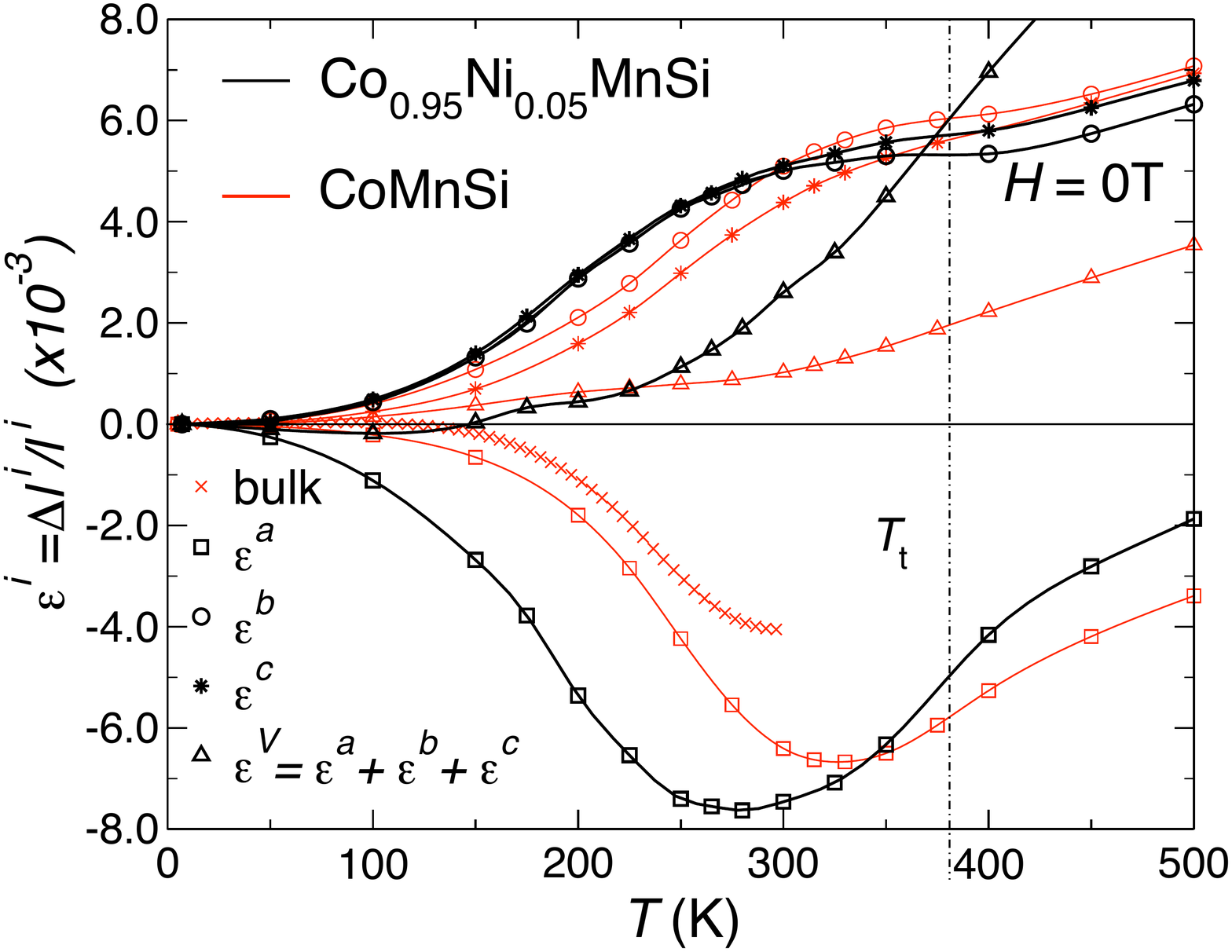}
\caption{(Color online) Thermal expansion $\epsilon^{i}$ of {\CMS} and {\CNMS} along each crystal axis measured on powders using HPRD. Lines are a guide to the eye. The largest change in cell parameters, including an $a$ axis NTE, occurs well below the metamagnetic transition temperature, {\Tt}.  Bulk thermal expansion of {\CNMS} measured using a dilatometer indicates that our as-prepared ingots are textured along the $a$-axis.
\label{fig:spont_ms}}
\end{figure}

In Figure~\ref{fig:spont_ms} we show a measure of the zero-field thermal expansion of the two samples involved in this study, {\CMS} and {\CNMS}, obtained using the HRPD neutron diffractometer. For CoMnSi, the sample was cooled down to 4.2~K and diffraction patterns were collected at 4.2~K, 50~K and then at 50~K intervals up to 500~K. Then the sample was cooled down to 225~K and additional patterns were collected at 225~K, 275~K, 315~K, 330~K and 375~K, resulting in 15 diffraction patterns. At each step the temperature was equilibrated for 20~min and a total neutron current of 75~$\mu$A was collected for each frame. Neutron diffraction by Co$_{0.95}$Ni$_{0.05}$MnSi was conducted as follows: at 4.2~K, 100~K and then each 50~K up to 300~K with a total neutron current of 75~$\mu$A and in a second experiment at other temperatures up to 500~K with 60~$\mu$A total neutron current. We used the GSAS code~\cite{larson_1994} to perform a Rietveld refinement of the crystal structure. From the resulting lattice parameters thermal expansion $\epsilon^{i}$, defined as ($l(T)-l(T_{0})/l(T_{0})$) was calculated, with $l(T)$ being one lattice parameter at temperature $T$ and $T_{0}$=5~K.

The N\'{e}el transition in zero magnetic field is second order, as evidenced by a broad peak in heat capacity measurements occurs at {\Tt}$\sim$380~K~\cite{sandeman_2006a}.   For CoMnSi we see that below {\Tt} there is a clear change in thermal expansion, with the most distinctive feature being a large negative thermal expansion (NTE) along the $a$-direction, as seen from the negative values of $\epsilon^{a}(T)$.  This NTE occurs between 4.2~K and around 330~K and amounts to a shrinkage of the $a$-axis of almost 0.7\%.   Weaker features of opposite sign were detected in the same temperature range along the $b$- and $c$-axis yielding anisotropic, positive thermal expansion of enhanced magnitude compared to a phonon-governed process. It should be noted that the crystal symmetry does not change over the whole measured temperature range.  The volume thermal expansion of a polycrystal can be estimated by $\epsilon^{a}+\epsilon^{b}+\epsilon^{c}$ and is also indicated in figure~\ref{fig:spont_ms}.  At around 250~K compensation of negative and positive thermal expansion leads to near-Invar-like temperature-independent behaviour of the calculated volume expansion.  Above {\Tt} the expansion of CoMnSi is more Debye-like.

The second alloy that we examined with the HRPD, Co$_{0.95}$Ni$_{0.05}$MnSi, also had a zero field {\Tt} $\sim$380~K and showed anisotropic thermal expansion similar to that of CoMnSi, but with the main features shifted to lower temperatures by about 70~K.  We may therefore infer some robustness to the anisotropic thermal expansion and $a$-axis NTE of CoMnSi-based helimagnets.


\begin{figure}
\includegraphics[width=\columnwidth]{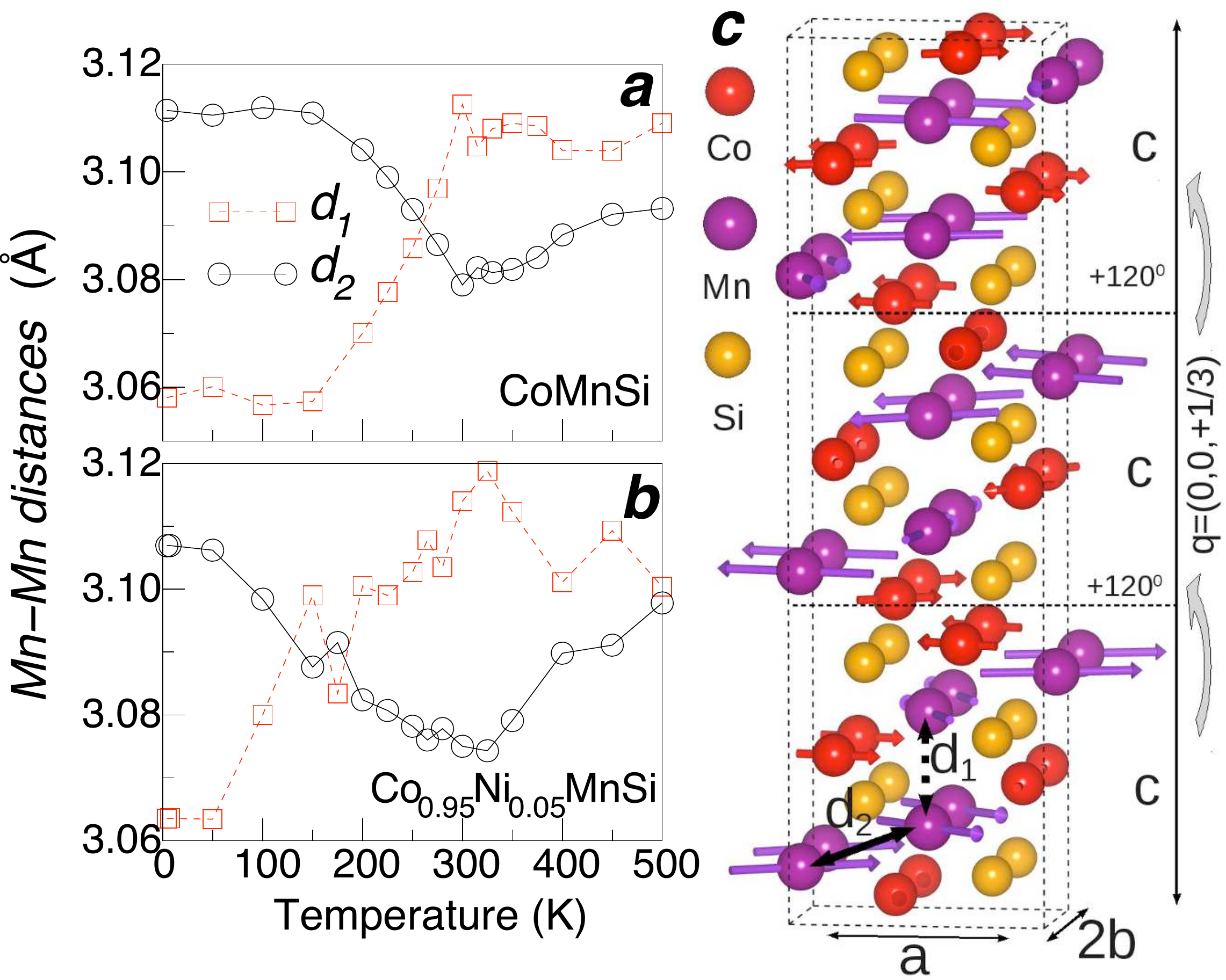}

\caption{(Color online) Temperature variation of Mn-Mn nearest-neighbour distances {\done} and  {\dtwo} in CoMnSi (a) and {\CNMS}, (b) from HRPD.  As with $a(T)$, the fastest variation of  {\done} and  {\dtwo} occurs well below {\Tt}.  The crystallographic positions of {\done} and  {\dtwo} are shown in (c), together with the closest commensurate version (for clarity, as modeled later) of the helical spin structure adopted by the Co and Mn atoms~\cite{niziol_1978a}.  3 unit cells are shown in the $c$ direction, drawn using VESTA~\cite{vestaref}.
\label{fig:d1d2}}
\end{figure}


\begin{figure}
\includegraphics[width=\columnwidth]{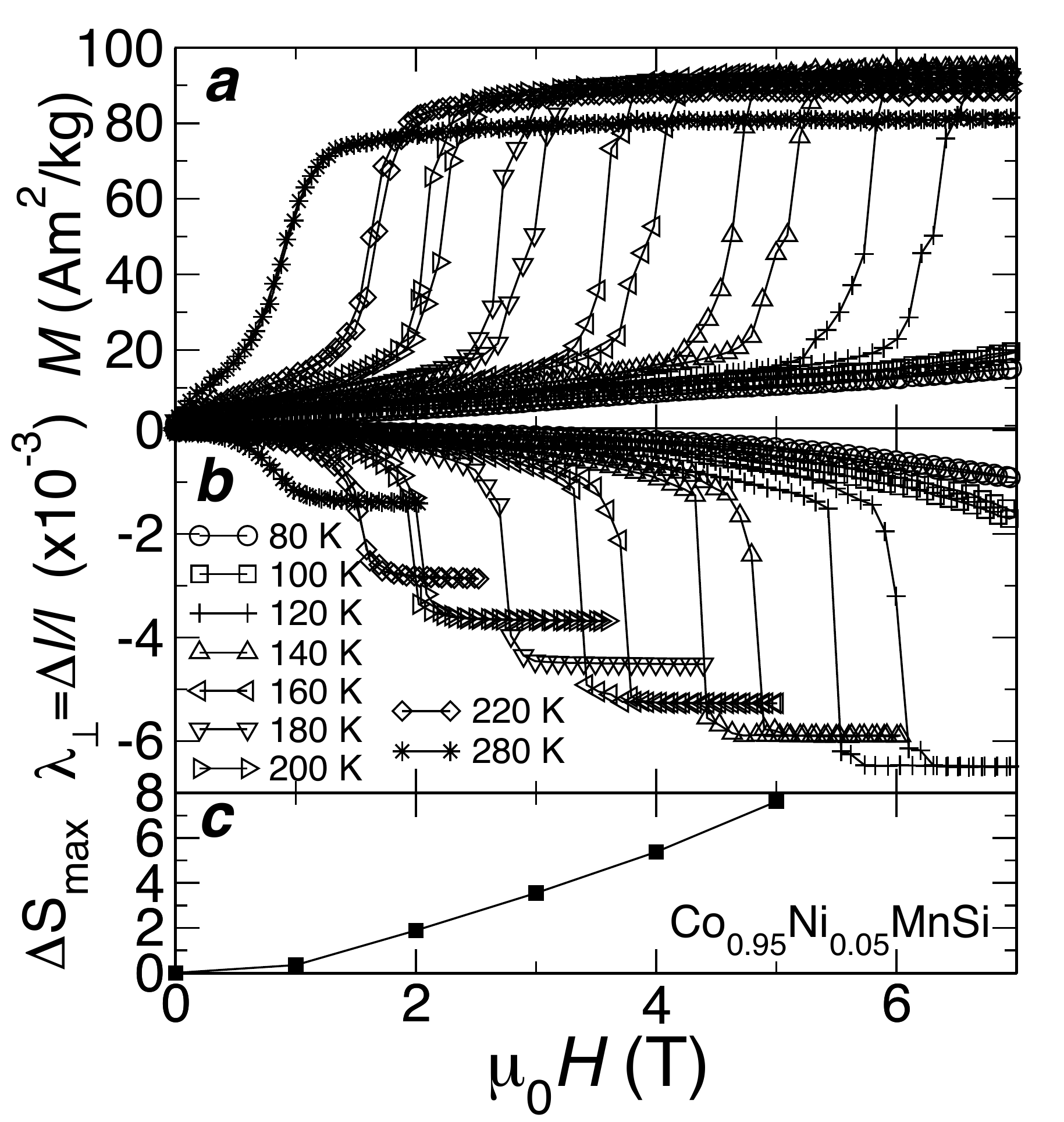}

\caption{(a) Isothermal magnetization of bulk Co$_{0.95}$Ni$_{0.05}$MnSi. (b) Transverse magnetostriction of a textured polycrystal of the same material. The magnetostriction is dominated by the $a$-axis texture of the sample (see Figure~\ref{fig:spont_ms}).  The onset of tricriticality is marked by an increase in both the magnetostriction and the peak MCE (c).
\label{fig:mag_ms}}

\end{figure}

We may also use the neutron diffraction data to track the temperature dependence of all interatomic separations.  The separation of localised Mn moments and the amount of intrinsic strain is believed to be very important to the observed magnetic properties of CoMnSi-based systems:  for CoMnSi either a N\'{e}el transition or (in quenched materials) a spontaneous transition to a high magnetisation state has previously been observed, with zero field literature transition values varying by up to 200~K.  We previously demonstrated a correlation between these values of {\Tt} and the value of the $a$-parameter found at 4.2~K, indicating the presence of a critical $a$ parameter for metamagnetism~\cite{sandeman_2006a}.  In this Letter we are able to demonstrate that the origin of such sensitive metamagnetism is in fact a giant magneto-elastic coupling, the largest yet seen in a metallic magnet.  In CoMnSi-based metamagnets, Mn carries the majority of the magnetic moment~\cite{niziol_1978a} and plays the most important role in determining magnetic properties.  In both {\CMS} and {\CNMS} we observe a crossover of the two nearest-neighbour Mn-Mn  separations, labelled  {\done} and {\dtwo} (see Figure~\ref{fig:d1d2}c) which occur in the same temperature range as the pronounced $a$-direction NTE, well below {\Tt}.  Both separations change, in opposite senses, by between 1\% and 2\%.    We may reasonably assume that such a drastic change in the local Mn environment acts as a precursor to the metamagnetic phase transition and that the critical $a$ parameter we found previously was a signature of this physics.  We leave the temperature dependence of the magnetic propogation vector to future studies as it lies outside the low-angle limit of HRPD.

We previously observed that in a magnetic field the N\'{e}el transition becomes a metamagnetic one to a high magnetisation state at temperature {\Tt}.  As the applied field is increased {\Tt} decreases sharply before going through a tricritical point where the onset of a finite latent heat and enhanced magnetocaloric effects are observed~\cite{morrison_2009a}.  We can observe the change in $a$ during this field-induced transition by measuring the forced magnetostriction of {\CNMS} samples.  We chose samples that were given an $a$-direction texture by the RF field during formation.  Such texturing is seen from a comparison of bulk thermal expansion data taken with the dilatometer with that from HRPD on powder (see Figure~\ref{fig:spont_ms}). In Figure~\ref{fig:mag_ms} we show both the magnetization and the magnetostriction of {\CNMS} as a function of applied field at different temperatures. Magnetostriction, is defined here for a polycrystal as $\lambda^{||,(\bot)}=(l(H)-l(H=0)/l(H=0))$, where $\lambda^{||}$ means the strain measured parallel to the applied magnetic field and $\lambda^{\bot}$ perpendicular to it.  We see the same tricritical point in both signals, as given by the onset of hysteresis.  Given the correlation between $a$-axis NTE and the giant change in {\done} and {\dtwo} we would expect that the field-induced change in {\done} and {\dtwo} also becomes first order at this point. Tricriticality occurs at a critical field of $\sim$2~Tesla, or at around 200~K and this temperature is relatively close to completion of the crossover in {\done} and {\dtwo} shown in Figure~\ref{fig:d1d2}.  We may thus assume that tricritiality occurs when the applied field drives {\Tt} low enough to couple to the giant underlying change in {\done} and {\dtwo}.  Field-dependent neutron studies are planned to explore this hypothesis.

Further evidence for such a conjecture is provided by using the above magnetisation measurements to calculate the MCE in both {\CMS} and {\CNMS} shown in figure~\ref{fig:mag_ms}(c).  We see an enhancement of peak isothermal entropy change associated with the metamagneti transition once the system goes through its tricritical point.  This feature again coincides with the temperature of maximal zero field variation of $a(T)$.  Entropy changes here are calculated from isothermal magnetisation curves using a Maxwell relation as in~\cite{sandeman_2006a} after confirming the reversibility of the metamagnetic phase transition.

\begin{figure}[b]
\includegraphics[width=\columnwidth]{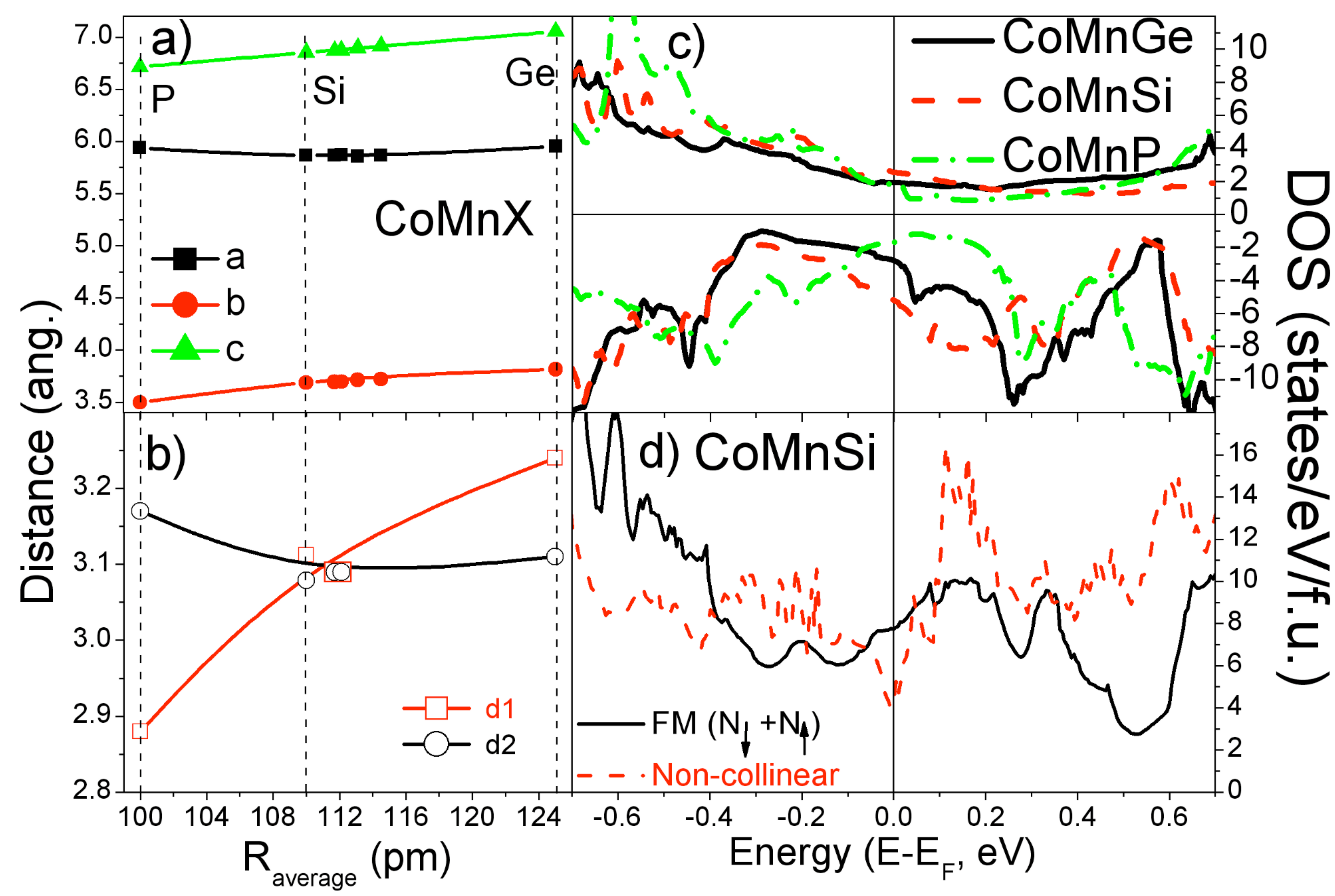}
\caption{(Color online) A comparison of CoMnX (X=P,Si,Ge). (a) and (b) show literature-derived lattice parameters and {\done},{\dtwo} at 300~K as a function of average p atom size~\cite{fruchart_1980a,sandeman_2006a,niziol_1980a,niziol_1982a}. In (c) we show the electronic densities of states (DOS) majority (upper figure) and minority (lower figure) spins near the Fermi energy, $E_{F}$ calculated for a ferromagnetic (FM) groundstate. In (d) we see that the calculated DOS at $E_{F}$ of non-collinear AFM CoMnSi is lower than that of a fictitious FM groundstate.}
\label{fig: vegard}
\end{figure}

We may explain our findings and their relevance by examining structural and magnetic data on other CoMnX materials in the same space group (X=P,Ge).   The detailed temperature variation of Mn-Mn separations is missing from previous studies, but for our purpose we may examine trends that link materials characterised at room temperature. In Figure~\ref{fig: vegard} we show the values of {\done},{\dtwo} and the unit cell parameters found in literature studies of CoMnP, CoMn(Si,Ge) and CoMnGe~\cite{fruchart_1980a,sandeman_2006a,niziol_1980a,niziol_1982a}.  On increasing the size of the p-block element, both the $b$ and $c$ parameter increase while the $a$ parameter decreases.  More notably we see a crossover in {\done} and {\dtwo}  that takes place at or near to the lattice dimension of CoMnSi, where the collinear ferromagnetism found in CoMnP and CoMnGe gives way to non-collinear antiferromagnetism.  It should be noted that {\done} has the stronger variation and may be the more important parameter. The plot shows that other materials in the space group, if engineered to give similar values of {\done} and {\dtwo}, may well give rise to the same competing exchange, giant magneto-elastic coupling and tricritical metamagnetism as found in CoMnSi.

The fact that exchange competition drives large magneto-elastic coupling is unusual.    We believe that the large magneto-elastic coupling found here is the result of a close competition between the relevant Mn-Mn and Mn-Co exchange interactions, and the energetic proximity of ferromagnetic, or low-$q$ spin configurations that correspond to quite different Mn-Mn separations.  To demonstrate the origin of the non-collinear AFM state and its energetic relation to ferromagnetic (FM) spin arrangements we have undertaken electronic structure calculations with the VASP code, based on density-functional theory (DFT)~\cite{vasp_ref}.  Details of the calculation are given elsewhere~\cite{gercsi_2010a}.  For a collinear FM state, we find that the calculated site-projected magnetic moments, together with the total magnetic moment per formula unit are fairly independent of the nature of the p-block atoms and are in accordance with the experimental findings by neutron diffraction in the cases of ferromagnetic CoMnP and CoMnGe~\cite{fujii_1979a,kaprzyk_1990a}. On the other hand, for CoMnSi a FM ground state gives $M_{\rm Mn}=2.88\mu_{\rm B}$ which overestimates the magnetic moment of the true non-collinear state ($M_{\rm Mn}\sim 2.2\mu_{\rm B}$~\cite{niziol_1978a}).  
Importantly we find no orbital contribution to the magnetic moment, and this confirms the simple picture of Mn being in a d$^5$ (Mn$^{2+}$) state giving rise to zero orbital angular momentum, if Hund's rule of maximum multiplicity is applied and each d-level is singly occupied.  We therefore believe the spin-orbit contribution to magneto-elastic coupling is much weaker than that due to exchange competition.  

The calculated DOS near the Fermi level of a collinear FM state are shown in Figure~\ref{fig: vegard} for CoMnX. A valley with an energy minimum at the Fermi level is observed for CoMnP in both minority and majority spins. The calculated total density of states ($N_{\rm tot}(E_{F})=N_{\uparrow}+N_{\downarrow}$) for CoMnP (1.7 states/eV/f.u.) and for CoMnGe (5 states/eV/f.u.) is considerably lower than 9 states/eV/f.u found for CoMnSi when forced to be ferromagnetic.  In the latter case, the high number of states makes ferromagnetism unstable and promotes a non-collinear AFM arrangement as was reported earlier in similar Mn-based orthorhombic systems~\cite{sredniawa_2001a, zach_2007a}.   This stability argument may be understood from either a Stoner viewpoint, whereby a magnetically ordered state is favoured by the reduction of DOS at the Fermi enegy, or as an extension of the picture in covalently bonded materials, where stability is found by maximally occupying the electronic bonding states.  We have found direct evidence for such a lowering of $N_{\rm tot}(E_{F})$ in the AFM case by also calculating the DOS of a commensurate (${\bf q}=(0,0,1/3)$) version of the helical magnetic structure found in CoMnSi (See Fig.~\ref{fig:d1d2} and Ref. ~\cite{niziol_1978a}).  The Mn states are less split by exchange in the non-collinear AFM state, as indicated by the partial density of states plots (PDOS, not shown here). The reduced exchange in this state shifts the steep valley seen in $N{}_{tot}$ at $\sim$0.5~eV to the Fermi level, stabilising non-collinear AFM, as seen in Figure~\ref{fig: vegard}.  The application of a field, or a change in temperature is sufficient, however, to bring about a change in magnetic structure and hence atomic separations.

In summary, we have shown that in materials with a composition close to {\CMS} there is giant magneto-elastic coupling evidenced by a crossover in nearest Mn-Mn separations.  In particular, the nearest neighbour separation can change by up to 2\%, away from the helimagnetic N\'{e}el temperature.  We have shown that this coupling is the precursor to a metamagnetic tricritical point with enhanced magnetostrictive and magnetocaloric effects.  While this is the largest magneto-elastic coupling seen in a metallic magnet, it should be possible to engineer materials of different compositions that have the same structural instability due to highly sensitive non-collinear AFM ordering. Future experiments will examine the magnetic field dependence of Mn-Mn separations and the magnetic structure.

We thank M. Avdeev, R.Bali, L.Cohen and K. Morrison for useful disccusions, K. Roberts for help with sample preparation and G.G. Lonzarich for use of sample synthesis facilities.  A.B. would like to thank EPSRC and Camfridge Ltd. for financial support.  K.G.S acknowledges financial support from The Royal Society.  The research leading to these results has received funding from the European Community's 7th Framework Programme under grant agreement 214864, as well as The Newton Trust and The Leverhulme Trust. Computing resources provided by Darwin HPC and Camgrid facilities at The University of Cambridge and the HPC Service at Imperial College London are gratefully acknowledged.

\end{document}